\documentstyle[12pt]{article}
\newcommand{\vtwo}{\vspace*{2mm}}
\newcommand{\htwo}{\hspace*{2mm}}
\newcommand{\honecm}{\hspace*{1cm}}

\newcommand{\comm}[2]{\left[\, #1 \, , \, #2 \, \right]}
\newcommand{\be}{\begin{equation}}
\newcommand{\bel}[1]{\begin{equation}\label{#1}}
\newcommand{\ee}{\end{equation}}
\newcommand{\bea}{\begin{eqnarray}}
\newcommand{\ba}{\begin{array}}
\newcommand{\eea}{\end{eqnarray}}
\newcommand{\ea}{\end{array}}

\newcommand{\ra}{\rightarrow}
\newcommand{\ru}{\rule[-2mm]{0mm}{8mm}}
\newcommand{\udl}{\underline}

\newcommand{\BEQ}{\begin{equation}}    
\newcommand{\BEA}{\begin{eqnarray}}
\newcommand{\EEQ}{\end{equation}}      
\newcommand{\EEA}{\end{eqnarray}}
\newcommand{\eps}{\epsilon}                      
\newcommand{\lmb}{\lambda}                       
\newcommand{\sig}{\sigma}                        
\newcommand{\rar}{\rightarrow}                   
\newcommand{\ket}[1]{\left|#1\right\rangle}      
\newcommand{\zeile}[1]{\vskip #1 \baselineskip}  

\newcommand{\build}[3]{\mathrel{\mathop{\kern 0pt#1}\limits_{#2}^{#3} }}

\catcode`\@=11
\def\numberbysection{\@addtoreset{equation}{section}
        \def\theequation{\thesection.\arabic{equation}}}
\numberbysection

\begin{document}
\baselineskip 0.3in
%
%
\begin{titlepage}
\null
\vskip 1cm
\begin{center}
\vskip 0.5in
{\Large \bf Schr\"odinger Invariance in Discrete Stochastic
Systems}
\vskip 0.5in
Malte Henkel and Gunter Sch\"utz
 \\[.3in]
{\em Department of Physics, Theoretical Physics, \\
University of Oxford, 1 Keble Road, Oxford OX1 3NP, UK}
\end{center}
\zeile{2}
%
\begin{abstract}
Local scale invariance for lattice models
is studied using new realizations of the Schr\"o\-din\-ger
algebra. The two-point function is calculated and it
turns out that the result can be reproduced from
exact two-point correlation functions
evaluated in the stationary state of several simple stochastic models.
\end{abstract}

\end{titlepage}

\newpage
%
%
\section{Introduction}
Among the most important concepts of physics is that of symmetries.
Particularly in the context of
critical phenomena, scale invariance and its consequences have been
thoroughly studied for a long time. More recently,
for isotropic, equilibrium critical points
in two dimensions, conformal invariance has allowed for significant
progress in the classification of the universality classes and the
elucidation of their properties, see \cite{ConRev} and references therein.

Much less is known if the system under study is  either strongly
anisotropic or time-dependent, or both. Such systems will at criticality
show an anisotropic scaling of, say, a correlation function $C(r,t)$
depending on a ``space'' coordinate $r$ (we take one space dimension
throughout) and ``time'' coordinate $t$
\BEQ
C(\lmb r, \lmb^{\theta}t)  = \lmb^{-2x/\theta} C(r,t)
\EEQ
where $\theta$ is the anisotropy or dynamical exponent
and $x$ is a scaling dimension. One may ask whether this strongly anisotropic
scale invariance can be sensibly generalized to include space-time
dependent rescaling functions $\lmb(r,t)$ instead of merely taking $\lmb$ to
be a constant. Indeed, at least for systems for which the equations of
motion reduce in the continuum limit to the free Schr\"odinger or
diffusion equation (and thus have $\theta=2$),
this seems to be possible. The group of scaling
transformations is then the Schr\"odinger group, which is the maximal
kinematic invariance group of the free Schr\"odinger equation \cite{Nied72}.
The Schr\"odinger group can be obtained as the non-relativistic limit of
the conformal group \cite{Baru73}.

The Lie algebra of the Schr\"odinger group in $d=1$ space dimension
is spanned from the set
$\{ X_{-1},X_{0},X_{1},Y_{-\frac{1}{2}},Y_{\frac{1}{2}},M_0 \}$
and the non-vanishing commutation relations are
\BEA \label{eq:SCHComm}
\left[ X_n , X_m \right] &=& (n-m) X_{n+m} \nonumber \\
\left[ X_n , Y_m \right] &=& \left( \frac{n}{2} - m\right) Y_{n+m} \\
\left[ Y_{\frac{1}{2}} , Y_{-\frac{1}{2}} \right] &=& M_0 \nonumber
\EEA
Note that $M_0$ commutes with the entire algebra and the Casimir operator
is \cite{Perr77}
\BEQ
Q= ( 4 M_0 X_0 - 2 \{ Y_{-\frac{1}{2}}, Y_{\frac{1}{2}} \} )^2
-2\{ 2 M_0 X_{-1}- Y_{-\frac{1}{2}}^{2},
     2 M_0 X_{1} - Y_{\frac{1}{2}}^2 \}
\EEQ
We shall denote the eigenvalues
of $M_0$ by ${-\cal M}$ and of $Q$ by ${\cal Q}$.
The projective unitary irreducible representations ({\sc puir}) of
(\ref{eq:SCHComm}) are classified \cite{Perr77}. There are four different
types of {\sc puir}s, but only one of them allows $M_0$ to be realized
non-trivially. Of these, the {\sc puir}s which have ${\cal Q}\neq 0$
give rise to an infinite set of internal quantum numbers
(spin does not arise in $d=1$). We consider the {\sc puir} with ${\cal Q}=0$
which is realized on
(scalar) wave functions $\psi(r,t)$ which are solutions of the Schr\"odinger
equation $(i\partial_t + 1/(2{\cal M}) \partial_r^2)\psi(r,t)=0$. It
takes the form $\psi \rar U\psi$, where \cite{Perr77}
\BEA
U \psi(r,t) &=& |-\gamma t+\alpha|^{-1/2}
\exp\left\{ -\frac{i {\cal M}}{-\gamma t+\alpha}\left( \frac{\gamma}{2} r^2
+\frac{\delta t -\beta}{2} v^2 - v r +(-\gamma t+\alpha) a v\right)\right\}
\nonumber \\
&\cdot& \psi\left( \frac{r-(-\gamma t+\alpha)a-(\delta t-\beta) a v}
{-\gamma t+\alpha},\frac{\delta t-\beta}{-\gamma t+\alpha} \right)
\label{eq:US}
\EEA
with $\alpha\delta-\beta\gamma=1$ and $\alpha,\beta,\gamma,\delta,v$ and $a$
are constants. This realization was found by Niederer \cite{Nied72} as
giving the maximal kinematic invariance group of the free Schr\"odinger
equation and, as shown by Hagen \cite{Hage72}, also arises when discussing
the Schr\"odinger invariance of non-relativistic free field theory.

The Lie algebra generators are readily found from (\ref{eq:US}) and read
\BEA
X_n &=& -t^{n+1}\partial_t -\frac{n+1}{2} t^n r\partial_r -\frac{n(n+1)}{4}
{\cal M} t^{n-1} r^2 \nonumber \\
Y_m &=& -t^{m+1/2}\partial_r -\left( m+\frac{1}{2}\right) {\cal M} t^{m-1/2} r
\nonumber \\
M_0 &=& -{\cal M} \label{eq:SGen}
\EEA
$X_{-1}$ and
$Y_{-\frac{1}{2}}$ generate translations in time and  space, respectively,
$X_0$ generates global scale transformation, $Y_{\frac{1}{2}}$ is the
generator of Galilei transformations and $X_1$ generates the special
Schr\"odinger transformations. In order to implement Galilei invariance of
the free Schr\"odinger equation, the scaling fields $\phi(r,t)$ of the
theory will under a Galilei transformation pick up a complex phase
characterized by the mass $\cal M$
and thus $\phi$ has necessarily to be complex \cite{Levy67}, as
is well known. Thus the field $\phi(r,t)$ is characterized by its scaling
dimension $x$ and its mass $\cal M$, while its complex conjugate
$\phi^*(r,t)$ is characterized by the pair $(x,-{\cal M})$.
This structure is to be kept when going, via the continuation
${\cal M}\rar (2i D)^{-1}$,
to a free diffusion equation with diffusion constant $D$.
The requirement of covariance of the two-point function under the
given Schr\"odinger transformation determines the two-point function
(written here in Euclidean time) completely \cite{Henk94}. One finds
\BEQ
<\phi_{1}(r_1,t_1) \phi^{*}_{2}(r_2,t_2)>= \Phi_0\,
\delta_{{\cal M}_1,{\cal M}_2}
\delta_{x_1,x_2} (t_1 -t_2)^{-x_1}
\exp \left[ - \frac{{\cal M}_1}{2} \frac{(r_1 -r_2)^2}{t_1-t_2} \right]
\EEQ
where $\Phi_0$ is a normalization constant. Note that the exponential factor
and the Bargmann  superselection rule ${\cal M}_1={\cal M}_2$ \cite{Barg54}
follow from covariance under the Galilei sub-group spanned by
$\{X_{-1},Y_{\pm\frac{1}{2}},M_0 \}$
alone. Several simple models can be shown to reproduce this result.
Three-point functions and plane surfaces have been considered as well
\cite{Henk94}.

These results for the correlation functions
do depend not so much on the algebraic structure but rather
on the particular realization of the Schr\"odinger algebra used. For
example, one could give another realization by using the space of
wave functions of the harmonic oscillator rather than free particles
as a carrier space \cite{Perr77,Nied73}.
Here, we want to consider the consequences of another
realization of (\ref{eq:SCHComm}). It differs from the free
particle realization defined above in that the carrier space is the space
of wave  functions which are periodic in momentum space. We shall define
this new realization in the next section and shall use it
to obtain the two-point function by demanding covariance.
Exact results for two-point functions
directly obtained from some simple lattice models coincide
with the expressions derived in section~2 and show that
this realization can be viewed as evidence for a Schr\"odinger
symmetry of lattice models.

\section{Lattice realization of the Schr\"odinger algebra}

We begin our discussion by noting that generators which satisfy the
Schr\"odinger algebra (\ref{eq:SCHComm}) can be constructed from the
generators $\{P,X,E,T,M\}$ of a much simpler
algebra ${\cal A}$ the generators
of which satisfy the relations
\bel{2-1}
\comm{P}{X} = \comm{E}{T} = 1
\ee
and
\bel{2-2}
\comm{A}{B} = 0
\ee
for all other pairs of generators $A,B \in \{P,X,E,T,M\}$. The generators
$X_{0,\pm1}, Y_{\pm \frac{1}{2}}, M_0$ defined as shown in table~1
\begin{table}
\begin{center}
\begin{tabular}{|c|l|} \hline
\ru generator & \\ \hline
\ru $X_{-1}$ & $- E$ \\
\ru $X_0$    & $- TE - \frac{1}{2} XP$  \\
\ru $X_1$    & $- T^2E - TXP - \frac{1}{2} MX^2$ \\
\ru $Y_{-\frac{1}{2}}$ & $ - P$ \\
\ru $Y_{\frac{1}{2}}$ & $ -TP - MX$  \\
\ru $M_0$ & $-M$ \\ \hline
\end{tabular}
\caption{ Realization of the Schr\"odinger
algebra in terms of generators of ${\cal A}$.}
\end{center}
\end{table}
satisfy indeed the commutation relations (\ref{eq:SCHComm}). Therefore
any representation of the algebra ${\cal A}$ gives rise to a representation
of the Schr\"odinger algebra. Clearly, $P = \partial_r$, $X=r$, $E=\partial_t$,
$T=t$ and $M={\cal M}$ (where $r,t,{\cal M}$ are numbers) satisfy relations
(\ref{2-1}) and (\ref{2-2}) and give rise to the "continuum" realization
(\ref{eq:SGen}) of the Schr\"odinger algebra (see also table~2).

The point we wish to make is that instead of choosing $P=\partial_r$,
one may represent $P$ by the symmetric difference operator
\bel{2-3}
P= \frac{2}{a}\sinh\left({\frac{a}{2}\partial_r}\right)
\ee
(which is defined by its expansion
in a power series). In order to preserve relations (\ref{2-1}) one may
now choose
\bel{2-4}
X= \frac{1}{\cosh{\frac{a}{2}\partial_r}} \, r \htwo .
\ee
The generators
$X_{0,\pm1}, Y_{\pm \frac{1}{2}}, M_0$ defined in table~1, but now given
by the expressions shown in table~2 still satisfy the commutation relations
(\ref{eq:SCHComm}).
\begin{table}
\begin{center}
\begin{tabular}{|c||l|l|} \hline
\ru generator & continuum & lattice \\ \hline
\ru $X_{-1}$ & $- \partial_t$ & $- \partial_t$ \\
\ru $X_0$    & $-t\partial_t - \frac{1}{2} r \partial_r$ &
 $-t\partial_t - \frac{1}{a \cosh(a/2 \, \partial_r )}
  r \sinh( a/2 \, \partial_r)$\\
\ru $X_1$    & $-t^2\partial_t - t r \partial_r - \frac{1}{2} {\cal M}r^2$ &
 $-t^2\partial_t-\frac{2 t}{a\cosh(a/2 \, \partial_r)} r
  \sinh(a/2 \, \partial_r)
 -\frac{{\cal M}}{2} \left(\frac{1}{\cosh(a/2 \, \partial_r)}r\right)^2 $\\
\ru $Y_{-\frac{1}{2}}$ & $-\partial_r $ &
 $-\frac{2}{a} \sinh( a/2 \, \partial_r)$  \\
\ru $Y_{\frac{1}{2}}$ & $-t\partial_r -{\cal M} r$  &
 $-\frac{2 t}{a} \sinh(a/2 \, \partial_r) -
 \frac{{\cal M}}{\cosh(a/2 \, \partial_r)} r$\\
\ru $M_0$ & $-{\cal M}$ & $-{\cal M}$ \\ \hline
\end{tabular}
\caption{The ``continuum'' and ``lattice'' realizations of the
Schr\"odinger algebra.}
\end{center}
\end{table}
The various operators are understood to be defined in terms
of a power series expansion.

Some intuitive understanding of the role of the parameter $a$ comes from
considering the generator
$Y_{-\frac{1}{2}}$ of ``translations''
\bel{2-5}
Y_{-\frac{1}{2}} f(r) = \left.\left.
-\frac{1}{a}\right(f(r+a/2)-f(r-a/2)\right) \htwo .
\ee
which, if we interpret $a$ as a lattice constant, may be understood as
a discretized symmetric lattice derivative operator.

The Casimir operator of the Galilei subalgebra has the abstract form
$C=ME-P^2/2$ and is therefore in the lattice realization given by
\bel{2-6}
C={\cal M} \partial_t - \frac{1}{a^2} \left(
\cosh{\left(a\partial_x\right)}-1\right) \htwo .
\ee
Thus $C$ can be viewed as the continuum or lattice Schr\"odinger operator
in the ``continuum'' or ``lattice'' realizations of table~2, respectively.
We emphasize that the generators given in table~2 do {\em not}
generate finite transformations but may still be viewed as generators of
infinitesimal transformations of some Lie group. In this paper, however,
we do not want to address the question of exponential maps of such generators.

We now turn to the two-point function
\BEQ
F = F(r_1,r_2;t_1,t_2) = <\phi_1 (r_1,t_1) \phi_2^{*}(r_2,t_2)>
\EEQ
and study  the consequences of covariance of $F$ under the
``lattice'' realization of (\ref{eq:SCHComm}).
We will use the short-hand
\BEQ
\partial_a = \frac{\partial}{\partial t_a} \;\; ; \;\;
D_a = \frac{\partial}{\partial r_a} \htwo .
\EEQ
In slight abuse of language
we shall furthermore refer to the action of the operators $Y_m$ and
$X_n$ as defined in table~2 as translation operator, Galilei
operator etc. Two-particle operators are defined as the
sum of two one-particle operators acting on the pair of coordinates
$(r_1,t_1)$ and $(r_2,t_2)$ respectively.

Demanding time translation
invariance implies $F=F(r_1,r_2;\tau)$ where $\tau=t_1-t_2$. Invariance
under $M_0$ gives the Bargmann superselection rule
${\cal M}_1={\cal M}_2={\cal M}$.
Translation invariance, generated by $Y_{-\frac{1}{2}}$, requires
\BEQ
\left[ \sinh\left(\frac{a}{2} D_1\right) + \sinh\left(\frac{a}{2} D_2
\right) \right] F(r_1,r_2;\tau)=0 \htwo .
\EEQ
This equation can be solved by
introducing the coordinates $R=r_1+r_2$ and $\rho=r_1 -r_2$ and writing
$F=f(R,\rho;\tau)$. We get
\BEQ
\cosh\left(\frac{a}{2}\partial_{\rho}\right)
\cdot \sinh\left(\frac{a}{2}\partial_{R}\right) f(R,\rho;\tau)=0 \htwo .
\EEQ
This has the solutions
\BEQ \label{eq:GBed}
f\left( R+\frac{a}{2},\rho;\tau\right) -
f\left( R-\frac{a}{2},\rho;\tau\right) = 0
\EEQ
and $f(R,\rho+a/2;\tau)=-f(R,\rho-a/2;\tau)$. Since we would like to recover
in the limit $a\rar 0$ the familiar continuum
solution, we have to choose (\ref{eq:GBed}). The second choice would lead to
$F(r_1,r_2;t_1,t_2)=0$ for $a\ra 0$ and we do not follow this possibility
any further.

Next, we consider Galilei invariance
\BEQ
\sum_{i=1}^{2} \left\{ \frac{2}{a}  t_i \sinh\left(\frac{a}{2} D_i\right)
+\eps_i {\cal M}_i \frac{1}{\cosh(a/2 D_i)} r_i \right\} F = 0
\EEQ
where $\eps_1=-\eps_2=1$. Using eq.~(\ref{eq:GBed})
and expanding the $\cosh$, we find
\BEQ
\frac{1}{\cosh(a/2 \partial_{\rho})}
\left[ \frac{\tau}{a} \sinh\left(a\partial_{\rho}\right) +
{\cal M}\rho \right] f\left(R+\frac{a}{2},\rho;\tau\right) = 0 \htwo .
\EEQ
Consider now the equation $(\cosh(a/2 \partial_{\rho}))^{-1} h(\rho)=0$.
We write $h$ in the form $h(\rho)=\sum_{k=0}^{\infty} h_{k}(\rho) a^k$.
Insertion then shows, after expanding the $\cosh$, that order by order
$h_k (\rho)=0$ for all $k$. We are thus left with
\BEQ \label{eq:LSG}
\left[ \frac{\tau}{a} \sinh\left(a\partial_{\rho}\right) +
{\cal M}\rho \right] f\left(R+\frac{a}{2},\rho;\tau\right) = 0
\EEQ
This can be solved via Fourier transformation with the result
\BEA
f\left(R+\frac{a}{2},\rho;\tau\right)&=& g\left(R+\frac{a}{2},\tau\right)
\frac{1}{2\pi} \int_{-\pi}^{\pi}
dk \, e^{(\tau/{\cal M}a^2) \cos k} e^{i k (\rho/a)} \nonumber \\
&=:& g\left(R+\frac{a}{2},\tau\right)
\Phi(\rho/a,\tau/{\cal M}a^2) \label{scale}
\EEA
If we interpret the $r_i$ indeed to denote the sites of a lattice,
$\rho/a=n$ is an integer and $\Phi(\rho/a,\tau/{\cal M}a^2) =
I_n(\tau/{\cal M}a^2)$, where $I_n$ is a modified Bessel function.
We also note the following properties of $\Phi$
\BEA
\partial_{\tau} \Phi(\rho/a,\tau/{\cal M}a^2) &=&
\frac{1}{{\cal M}a^2} \cosh(a \partial_{\rho}) \Phi(\rho/a,\tau/{\cal M}a^2)
\nonumber \\
\tanh\left(\frac{a}{2}\partial_{\rho}\right) \left( \rho
\Phi(\rho/a,\tau/{\cal M}a^2) \right) &=&
-\frac{2 \tau}{{\cal M}a} \sinh^2\left(\frac{a}{2}\partial_{\rho}\right)
\Phi(\rho/a,\tau/{\cal M}a^2) \label{eq:PhiId}
\EEA

We now consider the scale transformations generated by $X_0$. Scale
invariance requires that
\BEQ
\sum_{i=1}^{2} \left\{ t_i \partial_i + \frac{1}{a}
\frac{1}{\cosh(a/2 D_i)} r_i \sinh\left(\frac{a}{2} D_i\right) +
\frac{1}{2} x_i \right\} F = 0
\EEQ
where $x_i$ are the scaling dimensions of the fields $\phi_i$.
Following the same lines as before and using that
$[\rho,f(a\partial_{\rho})]=-a f'(a\partial_{\rho})$ for any
(differentiable) function $f$,
one obtains, with $x=(x_1+x_2)/2$
\BEQ
\left[ \tau\partial_{\tau}+\frac{1}{a}\tanh\left(\frac{a}{2}\partial_{\rho}
\right) \rho - \frac{1}{2} +  x \right] g(R,\tau)
\Phi(\rho/a,\tau/{\cal M}a^2) =0
\EEQ
Using (\ref{eq:PhiId}), this can be reduced to
\BEQ
\left[ \tau \partial_{\tau} + \frac{\tau}{{\cal M}a^2}
+ x-\frac{1}{2} \right] g(R,\tau) = 0
\EEQ
with the solution
\BEQ
g(R,\tau) = g_0(R) \tau^{-(x-1/2)}
\exp\left(-\frac{\tau}{{\cal M}a^2}\right)
\EEQ
where $g_0(R)$ satisfies $g_0 (R+a/2)= g_0 (R-a/2)$.
Finally, we demand invariance under the special Schr\"odinger transformation
generated by $X_1$:
\BEQ
\sum_{i=1}^{2} \left\{-t_i^2\partial_i-\frac{2 t_i}{a\cosh(a/2 D_i)}
r_i \sinh(a/2 D_i)
-\frac{\eps_i{\cal M}_i}{2} \left(\frac{1}{\cosh(a/2 D_i)}r_i\right)^2
- x_i t_i \right\} F = 0 \htwo .
\EEQ
A straightforward but tedious calculation shows that
the solution found so far is consistent, provided only that
\BEQ
x_1 = x_2 = x
\EEQ
The fields $\phi_1$ and $\phi_2$ must thus have the same scaling
dimension $x$ the value of which is not determined by
Schr\"odinger invariance. Summarising, the final result is
\bel{2-7}
F(n,t) \equiv F(r_1,r_2;t_1,t_2) = g_0 \,
\delta_{{\cal M}_1,{\cal M}_2} \delta_{x_1,x_2}
 t^{1/2 -x_1} e^{-t} I_n(t)
\ee
with
\BEQ\label{2-8}
t = \frac{t_1 - t_2}{{\cal M}_1 a^2} \honecm \mbox{and} \honecm
\;\; n = \frac{r_1-r_2}{a}
\EEQ
where we assumed that both $r_1$ and $r_2$ are integer multiples of the
``lattice'' constant $a$. As in continous space,
Schr\"odinger invariance on the lattice
determines the two-point function completely up
to the non-universal amplitude $g_0$,
the non-universal constant mass ${\cal M}$
and the universal critical exponent $x$. Note
that if we require $\phi_1$ and $\phi_2$ to be locally
conserved, i.e. $\partial_t \sum_n F(n,t) = 0$, then one has $x=1/2$.
If $r_1$ or $r_2$ are not integer multiples of $a$, the Bessel function
$I_n(t)$ has to be replaced by the scaling function (\ref{scale}). We can
consider $g_0$ as a normalization constant if both
$r_1$ and $r_2$ are integer multiples of $a$.

\section{Exact two-particle correlations in lattice models}

So far we have merely assumed the existence of physical systems
in which lattice Schr\"odinger invariance holds and studied some of
its consequences. Clearly it is desirable to have an example, even a
very simple one, where correlation functions as predicted in the
last section would be found. Such an example is many-particle lattice
diffusion.

\subsection{Symmetric, non-exclusive lattice diffusion}

Consider a system of arbitrarily many particles hopping stochastically
to their nearest neighbouring sites. We choose the probability of hopping
from site $i$ to site $i\pm1$ proportional to
the occupation number $n_i$ of the site $i$.
For this stochastic process one may either write a master equation
or, equivalently, a quantum Hamiltonian defining the time evolution
of the system \cite{doi,gr,adhr}. In this mapping
the state of the system at some time $t$ is given by (stochastic) occupation
numbers $\udl {n}= \{n_i\}$ where
$i$ is the number of a lattice site. Their dynamics
are given by a Schr\mbox{\"o}dinger equation for the probability distribution
$F(\udl{n},t)$ with a quantum Hamiltonian $H$
\bel{3-1}
\partial_t \ket{F(\udl{n},t)}= - H\ket{F(\udl{n},t)}
\ee
which for our very simple toy process is given by
\be
H = \frac{1}{2}
\sum_i (a_{i+1}^{\dagger} -a_i^{\dagger})(a_{i+1} - a_i) \htwo .
\ee
The operators $a_i^{\dagger}$ and $a_i$ satisfy bosonic commutation relations
$\comm{a_i}{a_j^{\dagger}}=\delta_{i,j}$. Each state in the system
is represented by a state vector $|\udl{n}\rangle$ which together with
the transposed vectors $\langle\udl{n}|$ form an orthonormal basis.
The creation operator $a_i^{\dagger}$ for the stochastic process is then
given by $\delta_{n,n-1}$ acting on site $i$, while the annihilation
operator $a_i$ is represented by $n\delta_{n,n+1}$ (here $\delta_{n,m}$ is the
Kronecker symbol).
In this notation one has $\ket{F(\udl{n},t)} = \sum_{\udl{n}} F(\udl{n},t)
|\udl{n}\rangle$. From the definition of the time evolution (\ref{3-1})
we find that a state vector at time $t$ is given in terms of the
vector at time $t=0$ by $\ket{F(\udl{n},t)}= \exp{(- Ht)}\ket{F(\udl{n},0)}$.
Time-dependent operators $A(t)$ (in euclidean time) are defined by
$A(t) = \exp{(Ht)} A \exp{(-Ht)}$.
Note that in one dimension this model is equivalent to an
interface growth model with unrestricted height gradients \cite{ans}.

$H$ has a degenerate ground state with energy $E_0=0$,
each sector with fixed particle number $N$ contains such a vector which
are the steady states of the system. Here
we shall study vacuum expectation values
\bel{3-2}
F(r_1,r_2;t_1,t_2) =
\langle 0 | a_{r_1}(t_1) a^{\dagger}_{r_2}(t_2) |0\rangle \htwo ,
\ee
i.e. correlation functions in the vacuum state containing no particles.
Here $r_{1,2}$ denote sites on the $1D$ lattice with lattice constant $a$.
We furthermore assume $t_1 \geq t_2$. In the language of stochastic
processes this is the conditional probability of finding the
particle on site $r_1$ at time $t_1$ provided it was on site $r_2$
at time $t_2$. Two-point correlation
functions in $N$-particle steady states can be reduced to
such correlators \cite{scsa}.

In order to compute the correlation function (\ref{3-2}) we first note
(a) that since we are considering a steady state correlator, $F$ depends
only on $r_1$, $r_2$ and
$\tau = t_1 - t_2$ and (b) that because of translational invariance,
$F$ depends only $r = r_1 - r_2$ and $\tau$.
Furthermore, operators $a^{\dagger}_i$ and $a_i$ satisfy the equations
of motion
\bel{3-3}
\ba{ccccc}
\partial_t a^{\dagger}_i & = & \comm{H}{a^{\dagger}_i} & = & -\frac{1}{2}
\left(a^{\dagger}_{i+1} + a^{\dagger}_{i-1}-2 a^{\dagger}_{i}\right) \vtwo \\
\partial_t a_i & = & \comm{H}{a_i} & = & ~~ \frac{1}{2}
\left(a_{i+1} + a_{i-1} - 2 a_{i}\right) \htwo .
\ea
\ee
This is equivalent to $C a^{\dagger}_i = C a_i = 0$ where $C$ is the
lattice Schr\"odinger operator (\ref{2-6}) with the eigenvalue
${\cal M}=1$ of the mass operator
$M$ defined by $\comm{N}{a_i} = - {\cal M} a_i = - a_i$ and
$\comm{N}{a^{\dagger}_i} ={\cal M} a^{\dagger}_i= a^{\dagger}_i$.

Eqs. (\ref{3-3}) are easy to integrate and one finds
\bel{3-5}
F(r,\tau) \equiv F(r_1,r_2;t_1,t_2) =
\mbox{e}^{-\tau} I_{r}(\tau)
\ee
which is in full agreement with the prediction (\ref{2-7}) with amplitude
$g_0 = 1$, mass ${\cal M}=1$, lattice constant $a=1$ and critical
exponent $x=1/2$ obtained from Schr\"odinger
invariance and current conservation which is implied in eqs. (\ref{3-3}).

\subsection{Driven non-exclusive lattice diffusion}

Symmetric lattice diffusion as discussed above appears to be simplest
many particle system which is invariant under the discretized Schr\"odinger
symmetry introduced in section~2. One may also study its driven version
where particles hop with different rates $p_R = q n_i$ and
$p_L= q^{-1} n_i$ from site $i$ to their right and
left nearest neighbouring sites $i\pm 1$ respectively.
In this case the Hamiltonian reads
\bel{3-6}
H = \frac{1}{2} \sum_{i} \left\{ q a^{\dagger}_{i+1}(a_{i+1} - a_i) +
    q^{-1} ( a^{\dagger}_{i+1} - a^{\dagger}_i ) a_{i+1} \right\}
\ee
and the correlation function (\ref{3-2}) is found by integrating the
equations of motion for $a_i$ and $a^{\dagger}_i$ to be
\bel{3-7}
F(r,\tau) \equiv F(r_1,r_2;t_1,t_2) =
\mbox{e}^{-b\tau-v r}
\mbox{e}^{-\tau} I_{r}(\tau)
\ee
with $v=-\ln{q}$ and $b=2\sinh^2(v/2)=\frac{1}{2}(q+q^{-1})-1$.
The same result can be obtained
from yet another lattice realization of the Schr\"odinger algebra. To see
this, we go back to the algebra ${\cal A}$ (\ref{2-1}). Rather than reusing
the realization given in eqs.~(\ref{2-3},\ref{2-4}), we now take
\BEQ
E=\partial_t + {\cal M}b \;\; , \;\; T=t \;\; , \;\;
P=\frac{2}{a}\sinh\left(\frac{a}{2}\left(\partial_r+{\cal M}v\right)\right)
\;\;,\;\; X=\frac{1}{\cosh(\frac{a}{2}(\partial_r +{\cal  M}v))} r
\EEQ
Using the definitions in table~1 this leads to a new
realization of the Schr\"odinger operators and one may go again through the
calculation of the two-point function. The result is
\BEQ \label{eq:NCLS}
F(n,t) = g_0\, \delta_{{\cal M}_1,{\cal M}_2} \delta_{x_1,x_2}
{\rm e}^{-{\cal M}_1 bt-{\cal M}_1 v n}t^{1/2-x_1} {\rm e}^{-t} I_{n}(t)
\EEQ
with $n$ and $t$ defined as in (\ref{2-8}).
This is reproduced by the result (\ref{3-7}) for asymmetric diffusion
with $g_0=1$, $a=1$, ${\cal M}=1$ and $x=1/2$.
Note that in the continuum limit $a \ra 0$ the correlation function
is proportional to
$\exp{(-{\cal M}v^2t/2 - {\cal M} v \rho)}\cdot
t^{-1/2} \exp{(-{\cal M}\rho^2/(2t))}$. This is the correlation
function of the undriven system, but observed
in a frame of reference moving with constant velocity $v$. On the lattice,
eq.~(\ref{eq:NCLS}) is understood as follows. A Galilei transformation will
bring us to the frame where particles in the mean are at rest. The effect
of this Galilei transformation produces the result (\ref{eq:NCLS}), and the
condition $b=2 a^{-2} \sinh^2(av/2)$ takes care that the stationary state has
indeed vanishing energy.

\subsection{Exclusive symmetric lattice diffusion}

We come back to lattice diffusion as defined in section~3.1 but demand that
on each site there can be at most one particle \cite{Ligget}. Then
the same expression (\ref{3-5})
for the two-point correlation function (\ref{3-2})
is found. The time evolution operator of this system is given by the spin 1/2
isotropic Heisenberg ferromagnet
\bel{3-8}
H = - \frac{1}{4} \sum_i \left( \sigma_i^x \sigma_{i+1}^x +
\sigma_i^y \sigma_{i+1}^y +
\sigma_i^z \sigma_{i+1}^z - 1\right)
\ee
where $\sigma_i^{x,y,z}$ are the Pauli matrices. This is an interesting
example of a process where the time evolution operator is given in terms
of operators satisfying a Hecke algebra \cite{Ritt94}.
The equations of motion for
the creation and annihiliation operators
$\sig^{\pm}=\frac{1}{2}(\sigma^x \pm i \sigma^y)$, respectively,
are nonlinear
\bel{3-9}
\partial_t \sig_i^+ = \frac{1}{2}
\left( \sig_{i+1}^+ + \sig_{i-1}^+ - 2 \sig_i^+ \right) +
 \left( \sig^{-}_{i+1}\sig^{+}_{i+1} + \sig^{-}_{i-1}\sig^{+}_{i-1}
\right) \sig_i^+
\ee
and a similar expression for $\sig_i^{-}$. However, the
nonlinear piece is proportional to the number operators at sites
$i\pm 1$ and therefore
vanishes in the expession (\ref{3-2}). The linear piece is again
the lattice Laplacian  which accounts for
the correlation functions being given by the same expressions as
in the non-exclusive case.

We only showed here that the two-point function can be reproduced
in the models studied and it is not immediately obvious that the
same should be true for the higher correlators.
But one might justify the consideration of
Schr\"odinger invariance in this context by noting that Schr\"odinger
invariance can be shown to hold also for at least certain types of
non-linear Schr\"odinger equations, e.g. $(i\partial_t +
\partial_r^2 + u |\psi|^2)\psi=0$, see \cite{Wei88,Fush93} for more
information. The equations of motion (\ref{3-9}) are a discretized
version of this equation. Furthermore, known results for a partially
integrated four-point function of the exclusion process \cite{ans},
suggest that one might predict
such correlators from an invariance of the kind discussed here, but
adapted to systems with boundaries.

\subsection{Other stochastic models}

The same type of correlation functions can be reproduced from other
systems as well. Consider the $1D$ kinetic Ising model with
Glauber dynamics \cite{Glau63}. The connected spin-spin
correlation function $<s_{r_1}(t_1) s_{r_2}(t_2)>_c$ of spins at the
lattice sites $r_{1,2}$ at times $t_{1,2}$, when
the system is in equilibrium at a temperature $T$, can be calculated
exactly. If in the initial
state there are no correlations between the spins, one has \cite{Glau63}
\BEQ
<s_{r_1}(t_1) s_{r_2}(t_2)>_c = e^{-\alpha t} I_{r}(\alpha \gamma t)
\EEQ
with  $r=r_1-r_2$, $t=|t_1-t_2|$, $\alpha$ is a constant reaction rate,
$\gamma=\tanh(2J/(k_B T))$ where $J$ is the Ising model coupling constant.
For $\gamma =1$ corresponding to the zero temperature static critical point
of the one-dimensional Ising model this is obviously in agreement
with the prediction eq.~(\ref{2-7}), upon
identification of the non-universal parameters. But also the
off-critical form of the correlation function ($\gamma\neq 1$)
is in agreement  with lattice Schr\"odinger
invariance (\ref{eq:NCLS}) with $v=0$, $x=1/2$ and choosing $b$ and $\cal M$
appropriately. This corresponds to the invariance group of a Schr\"odinger
equation in a constant potential, proportional to $b$.

Another example is provided by the stochastic adsorption-desorption
process of dimers \cite{Gryn94}. Consider a $1D$ lattice whose sites
may be empty or occupied by at most one particle. A pair of
empty nearest-neighbour sites may become occupied with rate $\eps$
(dimer adsorption), and a pair of occupied nearest-neighbor sites
may become empty with the same rate $\eps$ (dimer desorption). A single
particle may hop from an occupied site to an empty nearest neighbor site
with a rate $h$. Let $n_r(t)$ denote the occupation number of site $r$
at time $t$. It can be shown that the system evolves  towards a steady
state, characterized by $<n_r>=1/2$. In the steady state, the density-density
correlation is on an infinite lattice \cite{Gryn94}
\BEQ
<n_{r_1}(t) n_{r_2}(0)> =\frac{1}{4} e^{-2(\eps+h)t} I_{r}(2(h-\eps)t)
\EEQ
with $r=r_1-r_2$. This correlation again reproduces lattice
Schr\"odinger invariance (\ref{eq:NCLS}), when choosing $b$ appropriately.

\section{Conclusions}

We have introduced new realizations of the Schr\"odinger algebra
as describing dynamical symmetries of discrete lattice systems.
The two-point correlation function is found from the requirement
that it covariantly transforms under these realizations and includes
the previously known result of the continuum as the special
limiting case $a\rar 0$.
We find that the two-point function so obtained can be reproduced
from the two-particle correlation function, calculated in the
stationary state, of several simple stochastic models.

The construction of lattice realizations of continuous space-time
symmetries using the algebra (\ref{2-1}), (\ref{2-2}) extends beyond
the few cases studied here. Likewise, one may generalize this approach
to higher dimensions. This will be reported elsewhere.

\zeile{2}
\noindent{\bf Acknowledgements}
\zeile{1.5}
\noindent This paper is dedicated to V. Rittenberg on the
occasion of his $60^{\rm th}$ birthday. It is a pleasure to thank him
for generously sharing his time and insight in many stimulating discussions.
G.S. thanks the Isaac Newton Institue, cambridge, where part of this
work was done for kind hospitality.
This work was supported by the SERC.

\end{document}